\pdfoutput=1
\documentclass[a4paper,fleqn,usenatbib]{mnras}

\usepackage{tikz}
\usetikzlibrary{patterns}
\usetikzlibrary{positioning}

\usepackage{blindtext}
\usepackage{mathdots}
\usepackage{enumitem}
\usepackage[ruled]{algorithm2e}

\usepackage{pgfplots}
\usepackage{subfig}

\usepackage[T1]{fontenc}
\usepackage{ae,aecompl}

\usepackage{graphicx}	
\usepackage{amsmath}	
\usepackage{amssymb}	

\newcommand{\HII}{\textsc{H\,ii}}

\title[Persistent topology of reionisation]{Persistent topology of the reionisation bubble network. I: Formalism \& Phenomenology}

\author[Elbers \& van de Weygaert]{Willem Elbers$^{1}$ and Rien van de Weygaert$^{1}$
\\
$^{1}$Kapteyn Astronomical Institute, University of Groningen, P.O. Box 800, 9700AV Groningen, the Netherlands}

\date{Last updated 2 December, 2018; in original form 2 December, 2018}

\pubyear{2015}

\pubyear{2018}

\begin{document}
\label{firstpage}
\pagerange{\pageref{firstpage}--\pageref{lastpage}}
\maketitle

\begin{abstract}
We present a new formalism for studying the topology of $\HII$ regions during the Epoch of Reionisation, based on persistent homology theory. With persistent homology, it is possible to follow the evolution of topological features over time. We introduce the notion of a persistence field as a statistical summary of persistence data and we show how these fields can be used to identify different stages of reionisation. We identify two new stages common to all bubble ionisation scenarios. Following an initial pre-overlap and subsequent overlap stage, the topology is first dominated by neutral filaments (filament stage) and then by enclosed patches of neutral hydrogen undergoing outside-in ionisation (patch stage). We study how these stages are affected by the degree of galaxy clustering. We also show how persistence fields can be used to study other properties of the ionisation topology, such as the bubble size distribution and the fractal-like topology of the largest ionised region.
\end{abstract}




\section{Introduction}

The Epoch of Reionisation was a cosmic phase transition in which the neutral hydrogen of the post-recombination era was ionised by the first luminous objects. Reionisation coincides with and influences the formation of the first galaxies, resulting in a complex and non-linearly evolving ionisation fraction field $x_\text{II}=N_\text{II}/(N_\text{I}+N_\text{II})$. The topology of this ionisation field has been the subject of sustained theoretical interest. One hope is that the topology will tell us about the physical processes involved and in particular about the sources responsible for reionisation \citep{friedrich11,katz18}. With currently ongoing observations of the redshifted 21-cm line \citep{beardsley16,patil17,kerrigan18}, we will for the first time gain access to statistics of the 21-cm field and the closely related ionisation field. If techniques improve sufficiently, we will even be able to image the ionisation field through 21-cm tomography, which is one of the goals of the Square Kilometre Array \citep{mellema15}. To connect these observations to the many simulations\footnote{State of the art simulations include \citet{gnedin14,iliev14,ocvirk16,pawlik16,doussot17}. Semi-numerical approximations are also commonly used \citep{mesinger07,choudhury09,mesinger11,zahn11,majumdar14,hutter18}.} of the reionisation era, it is important to develop robust measures that capture a sufficient level of detail of the ionisation topology and are appropriate for every stage of the ionisation process. This study is an effort to develop such a measure by borrowing from the theory of persistent homology. In this first paper of two, we explain our methodology and illustrate the usefulness of persistent homology with a number of phenomenological models. In a follow up paper, we apply these ideas to more realistic scenarios.

\subsection{Topology of reionisation}

An early qualitative description of the topology of reionisation goes back to \citet{gnedin00}, who identified three stages of reionisation. During the \emph{pre-overlap stage}, radiation emitted by the first luminous objects ionises the dense surrounding gas, forming localised bubbles of ionised material. These bubbles then expand into the low-density intergalactic medium. In a second \emph{overlap stage}, the ionised regions merge and the global ionisation fraction rises rapidly. Finally, in the \emph{post-overlap stage}, the remaining high-density neutral pockets are ionised from the outside. This picture of reionisation can be described in terms of \emph{inside-out} and \emph{outside-in} reionisation \citep{lee08,choudhury09,friedrich11}. These terms refer to the ionisation of high-density regions: high-density regions containing ionising sources are ionised first and their bubbles expand outward (inside-out), but high-density regions without ionising sources are ionised from the outside at the end of reionisation (outside-in). Rather than high-density pockets, high-density filaments might also be the last regions to be ionised \citep{finlator09}. Either way, the degree to which outside-in reionisation occurs depends on the minimum halo mass necessary for ionising sources to form, demonstrating one way in which the topology reflects the underlying physics. Another example is the degree of galaxy clustering, which affects the patchiness of the ionisation field \citep{iliev14}.


\begin{figure*}
  \centering
  \subfloat[$\alpha=0.7$] {
    \centering\includegraphics[width=.3\linewidth]{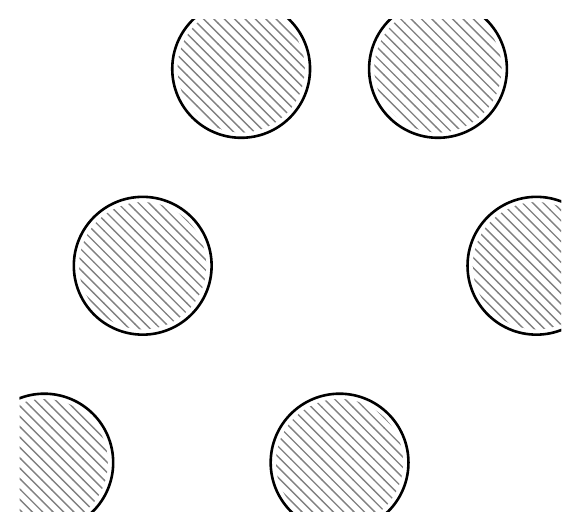}
  }
  \subfloat[$\alpha=1.2$] {
    \centering\includegraphics[width=.3\linewidth]{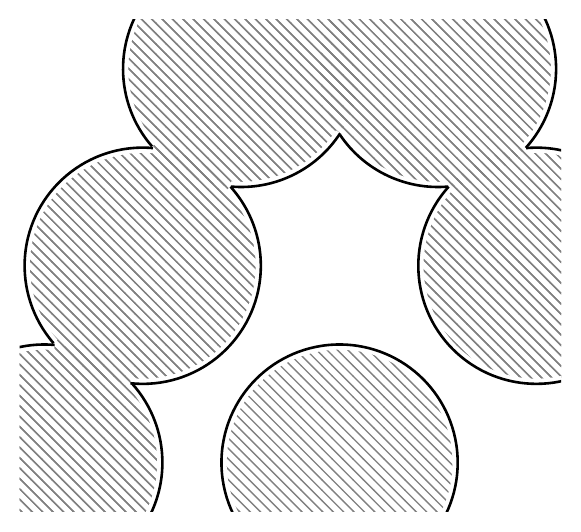}
  }
  \subfloat[$\alpha=1.5$] {
    \centering\includegraphics[width=.3\linewidth]{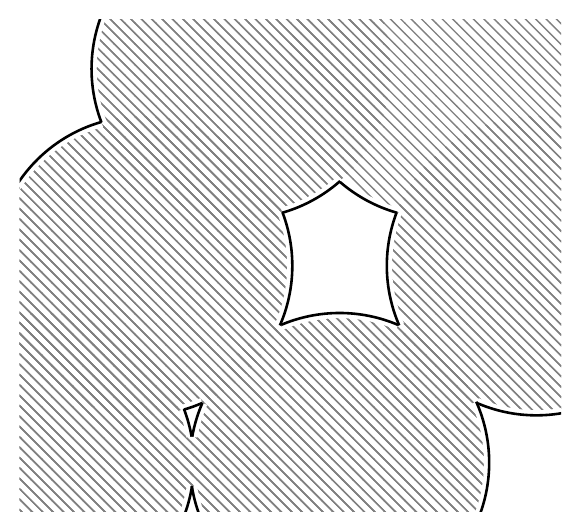}
  }
  \caption{Filtration of a uniformly sized bubble network.}
  \label{fig:bubble_net}
\end{figure*}

The reionisation process has been most commonly quantified with the 21-cm power spectrum or more directly with the power spectrum of the ionisation field, which constitutes the dominant component of the 21-cm power spectrum during the latter half of reionisation \citep{iliev14}. The 21-cm power spectrum is the first observable that is likely to be measured and contains valuable information. The overall amplitude of the 21-cm power spectrum tracks the progress of reionisation, since the differential brightness temperature is proportional to the fraction of neutral hydrogen\footnote{Indeed, we have \citep{pritchard12}:
\begin{align*}
\delta T_{21}(z) = T_0(z)(1+\delta_b)(1-x_\text{II})\left(1-\frac{T_\text{CMB}(z)}{T_\text{S}}\right),
\end{align*}

\noindent
where $T_\text{S}$ is the spin temperature, $T_0(z)$ a function of cosmological parameters and redshift $z$, and $\delta_b$ the baryonic overdensiy.}. The amplitude of the ionisation power spectrum peaks in the middle of reionisation when variance in the ionisation field is highest \citep{hutter18}. A general finding is that once reionisation has started, the ionisation power spectrum peaks at some scale indicative of a characteristic bubble size, which increases as reionisation progresses and bubbles merge \citep{furlanetto4b,zahn11,majumdar14,iliev14,hong14,dixon15,hutter18}. The power spectrum can also be used to identify more complex patterns in the ionisation topology. \citet{friedrich11} found two peaks and explained this with two periods of ionisation bubble formation interceded by a period of suppresion. The slope of the power spectrum may indicate to what degree ionisation occurred outside-in \citep{choudhury09}. Finally, the 21-cm power spectrum also carries information on pre-reionisation physics \citep{mesinger11}. Nevertheless, the power spectrum is not enough to characterise the evidently non-Gaussian ionisation field. \citet{kakiichi17} nicely demonstrated that the 21-cm signal from a radiative transfer simulation is morphologically very different from a Gaussian random field with the same power spectrum. Hence, complementary observables such as the bispectrum \citep{shimabukuro17} are needed (this paper introduces another such observable).

A common alternative has been to study the morphology of individual ionisation bubbles. Many authors have looked at the size distribution of ionisation bubbles \citep{furlanetto04,mcquinn07,mesinger07,friedrich11,zahn11,malloy13,lin16,kakiichi17,giri17} or at the shape of such bubbles \citep{gleser06,iliev06,furlanetto16a,bag18}. They typically find that the bubble radius is approximately log-normally distributed with a characteristic scale that increases and a variance that decreases as reionisation progresses. 
 
Recently, reionisation has also been fruitfully studied from the perspective of percolation theory \citep{furlanetto16a,bag18}. A salient feature of the qualitative description above is the sharp rise in ionisation fraction during the overlap stage. This can be understood as a phase transition associated with the percolation of ionisation bubbles. Near the phase transition, the ionised regions demonstrate the scaling behaviour expected from universality.
 
In terms of purely topological measures, the most basic is probably the number $k$ of connected components. Starting from a discretised snapshot of the ionisation field, this can be determined by applying a friends-of-friends algorithm \citep{friedrich11}, watershed algorithm \citep{platen07,lin16} or technique called granulometry \citep{kakiichi17} to the points with an ionisation fraction above a certain threshold. The evolution of the number of ionised regions alone can already tell the qualitative story of emerging and then rapidly merging bubbles. A more detailed variation is to follow the evolution of individual ionised regions and to construct merger-trees, which allows one to study the number density of new, expanding, and merging regions over time \citep{chardin12}. 

Another elementary topological property is the genus $g$, which is the number of cuts one can make without increasing the number of components, or the related Euler characteristic $\chi=2k-2g$. More complex still, the Minkowski functionals combine geometric properties such as the volume, surface area, and mean curvature of the ionised region with the Euler characteristic. Both genus and Minkowski functionals have been applied in this context. Different stages of reionisation can be distinguished by means of genus curves \citep{lee08} and Minkowski functionals \citep{gleser06}. Both can be used to constrain various source properties \citep{friedrich11}. The 21-cm field too has been studied with genus curves \citep{hong14} and Minkowski functionals \citep{yoshiura16}, both agreeing that they can be used to constrain physics if accurate images of the 21-cm signal were available. \citet{bag18} used ratios of Minkowski functionals called \emph{shapefinders} \citep{sheth03,shandarin04} to express such properties as the length, thickness, and breadth of the largest ionised region. They found that the largest ionised region that emerges during the phase transition has a complex and highly filamentary topology.

To summarise, most studies thus far have focused on global features of the topology such as the Euler characteristic or on the morphology of individual ionisation bubbles. However, the picture of disconnected ionisation bubbles is only appropriate during the pre-overlap stage when the global ionisation fraction is relatively small. During most of reionisation, most of the ionised material is contained in one connected structure that stretches the length of the Universe and has a complicated and fractal-like topology \citep{furlanetto16a,bag18}. We would therefore like to find tools that help us understand the topology during the later stages of reionisation, especially since the later stages are easiest to observe through the 21-cm signal. 

\subsection{Persistent homology}

\begin{figure}
	\centering\includegraphics[width=\linewidth]{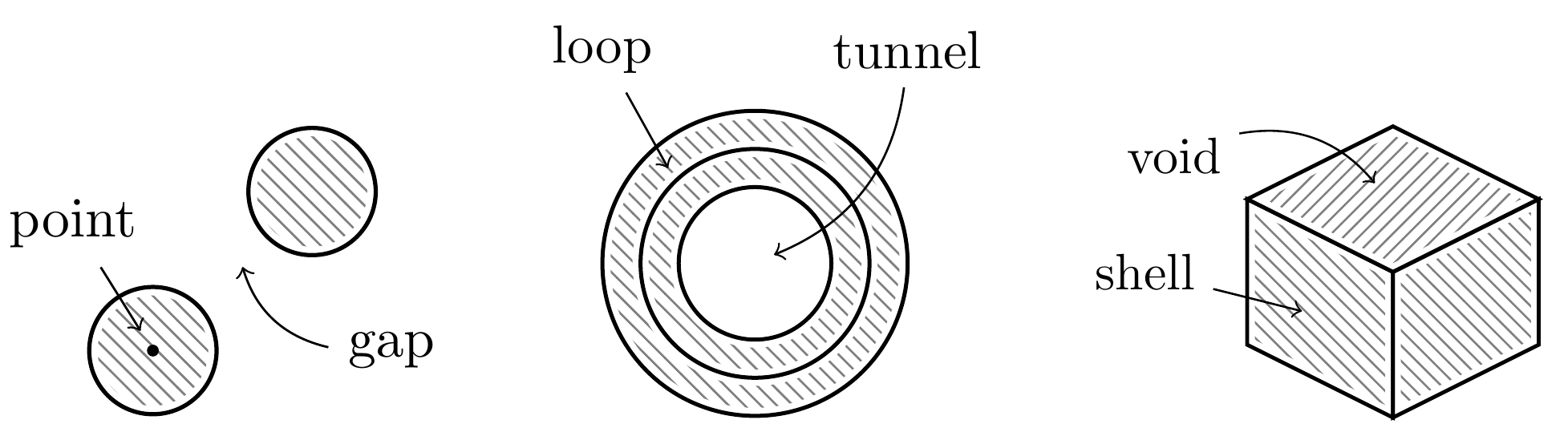}
	\caption{The homology of an object refers to the distinct classes of loops that can be drawn on it, or equivalently about its boundaries and holes. A $k$-dimensional loop (point, loop, shell) can be continuously deformed until it meets a $k$-dimensional hole (gap, tunnel, void). Shown are $k=0,1,2$.}
	\label{fig:holes}
\end{figure}

In this paper, we show that persistent homology is ideally suited to study the process of cosmic reionisation through its topology. As a subfield of mathematics, topology is concerned with properties that are preserved under continuous deformations (like bending or stretching). An important example of such a property is the number of holes. Counting holes is therefore a useful way to distinguish topologies. In figure \ref{fig:holes}, we see three examples of holes in different dimensions. A 0-dimensional hole is a gap that separates a connected component, like a distinct $\HII$ region, from the space surrounding it. There is one gap for each component, so we often blur the distinction. A 1-dimensional hole is an opening like the cross section of a tunnel. Finally, a 2-dimensional hole is a cavity or void surrounded by a shell. We will generally refer to gaps, tunnels, and voids as \emph{topological features}.

We are interested in topological features in the ionisation field. The connected components of this field are simply the ionisation bubbles or the distinct $\HII$ regions. The tunnels are neutral filaments that pierce through the ionisation bubble network. Voids are patches of neutral hydrogen enclosed by ionised material. We stress that these are voids in the ionisation field, often corresponding to overdensities and distinct from cosmological voids, which correspond to underdensities. We refer to these features as ionised bubbles $(k=0)$, neutral filaments or tunnels $(k=1)$, and neutral patches $(k=2)$.

In homology theory, we count holes by classifying the loops that can be drawn on an object. This is possible because of a correspondence between loops and holes. We are primarily interested in the so-called Betti numbers. Formally, the $k$th Betti number $\beta_k$ is the rank of the $k$th homology group, which contains the distinct classes of $k$-dimensional loops. Intuitively, the $k$th Betti number is simply the number of $k$-dimensional holes. In other words, the zeroth Betti number $\beta_0$ describes the number of connected components, the first Betti number $\beta_1$ the number of tunnels, and the second Betti number $\beta_2$ the number of voids. 

Together, the Betti numbers contain strictly more information than the Euler characteristic $\chi = \beta_0 - \beta_1 + \beta_2$. As the number of ionised regions is initially much larger than the number of enclosed filaments and neutral patches, the Euler characteristic has sometimes been thought of as a measure of the number of bubbles: $\chi\approx\beta_0$. However, it is interesting to consider the Betti numbers separately. We should for instance be able to see the filamentary nature of reionisation by looking at $\beta_1$. Neutral patches undergoing outside-in ionisation can be identified by looking at $\beta_2$.

We can go one step further by taking topological persistence into account \citep{edelsbrunner00,zomorodian05}. Rather than dealing with a static object, we consider a nested sequence of objects\footnote{In which each element of the sequence contains the previous element.} called a filtration. It facilitates a formal mathematical description of the hierarchical evolution of structure. Intuitively, we picture a filtration as an expanding bubble network, as depicted in figure \ref{fig:bubble_net}. Each element in the sequence is assigned a scale $\alpha$. By studying the topology at every scale, we compute a birth date $\alpha_\text{birth}$ and death date $\alpha_\text{death}$ for all topological features. In figure \ref{fig:bubble_net}, we see the death of multiple gaps and the birth of two tunnels. The difference $\alpha_\text{death}-\alpha_\text{birth}$ is the \emph{persistence} of a feature. In a persistence diagram, all features are plotted in the $(\alpha_\text{birth},\alpha_\text{death})$-plane. Persistence diagrams contain even more information than Betti numbers, which only count the numbers of topological features at a given scale. For example, if we consider the filtration of a bubble network along the time axis, we can see not just the number of neutral patches but also how long it takes for them to be ionised.

\begin{figure*}
  \centering
  \subfloat[Voronoi tesselation] {
    \centering\includegraphics[width=.24\linewidth]{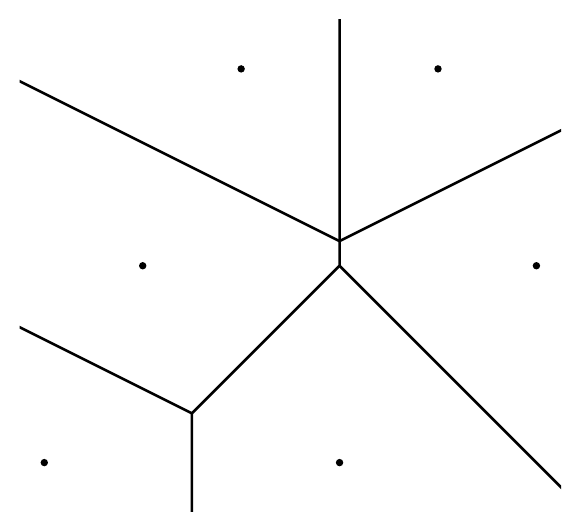}
  }
  \subfloat[Delaunay triangulation] {
    \centering\includegraphics[width=.24\linewidth]{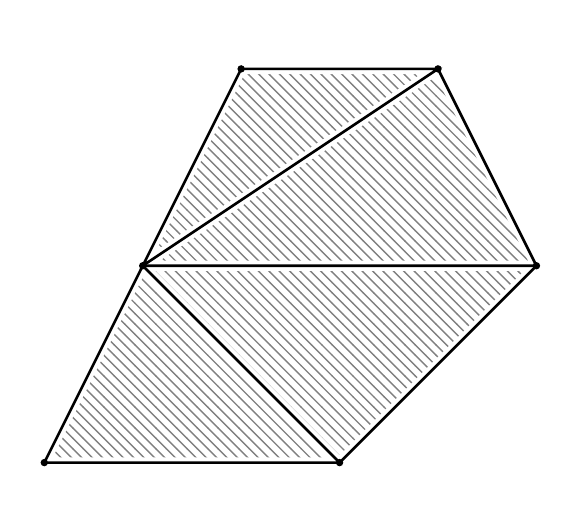}
  }
  \subfloat[Alpha-shape for $\alpha=1.2$] {
    \centering\includegraphics[width=.24\linewidth]{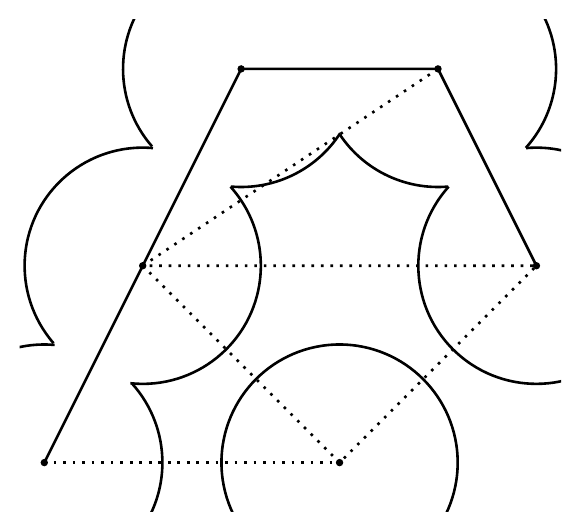}
  }
  \subfloat[Alpha-shape for $\alpha=1.5$] {
    \centering\includegraphics[width=.24\linewidth]{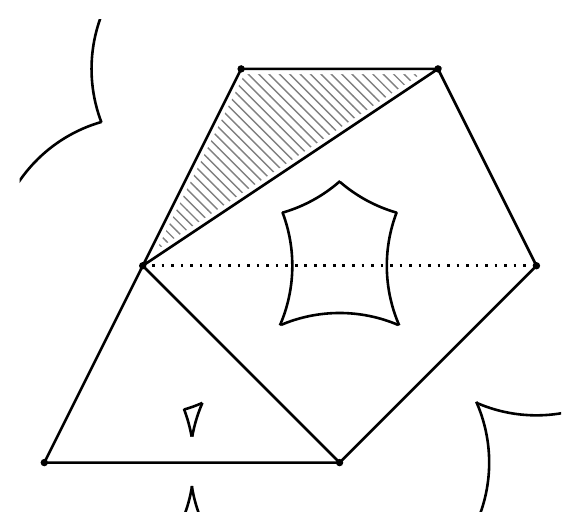}
  }
  \caption{The idea behind simplicial homology is that we can study the homology of a complex object by looking at the homology of an associated structure of points, lines, and triangles (simplices), which are computationally easier to handle. In the final panel, notice that the bottom triangle is not filled in, correctly capturing the opening that exists in the bubble network. Compare figure \ref{fig:bubble_net}.}
  \label{fig:alpha_shape}
\end{figure*}

In the context of reionisation, there are three interesting dimensions along which to build a filtration.

\begin{enumerate}[label=(\roman*), wide=0pt, widest=99,leftmargin=\parindent, labelsep=*]
	\item \emph{Time.} The most straightforward interpretation is to imagine the filtration as a bubble network evolving over time. In this case, $\alpha_\text{birth}$ and $\alpha_\text{death}$ are literally the birth and death dates of topological features. The persistence is simply the lifetime of a feature. A temporal filtration shows the hierarchical build-up of structure.
	\item \emph{Space.} Given a time slice of the ionisation history, we can also probe the connectivity structure of the bubble network. In this case, $\alpha_\text{birth}$ and $\alpha_\text{death}$ refer to spatial scales at which features arise. The persistence is now a measure of the topological significance of a feature. A spatial filtration looks into the multi-scale structure that emerges as a result of hierarchical evolution.	  
	\item \emph{Ionisation fraction.} In this paper, we assume a binary ionisation field. However, we can also construct a filtration by lowering the ionisation threshold (the ionisation fraction above which a point is considered ionised). The persistence of a feature is now interpreted as the differential ionisation fraction of the hole. For instance, the persistence of an opening tells us about the ionisation state of the enclosed filament. In this study, we consider only filtrations along the first two dimensions.
\end{enumerate}

\noindent
With developments in computational topology over the past two decades, persistent homology is now readily applicable in various practical situations. It has become the preeminent tool of topological data analysis \citep{zomorodian12,wasserman18}. In cosmology, persistent homology has previously been applied to the cosmic web \citep{weygaert11,sousbie11,nevenzeel13,pranav16,xu18}, to Gaussian random fields \citep{feldbrugge12,park13,cole18,feldbrugge18,pranav18b}, and to interstellar magnetic fields \citep{makarenko18}.

We further discuss the theory of filtrations and homology in section \ref{sec:theory}. In section \ref{sec:filtrations}, we describe our methodology and elaborate on the interpretation of bubble network filtrations. In section \ref{sec:sources}, we discuss how the bubble network depends on the properties and spatial distribution of ionising sources. The interpretation of persistent homology is explained in section \ref{sec:results} using a number of phenomenological models. Finally, we conclude in section \ref{sec:conclusion}.

\section{Theory}\label{sec:theory}

Our formalism makes use of persistent homology theory to analyse bubble networks. We also borrow a tool from computational topology called $\alpha$-shapes to model these bubble networks. The first part of this section deals with $\alpha$-shapes and its weighted generalisation. The latter is needed to model non-uniform bubble networks. We discuss an alternative to $\alpha$-shape filtrations in section \ref{sec:field_filtrations}. The rest of the section is concerned with homology theory, topological persistence, and the statistics of persistence diagrams.

\subsection{Alpha-shapes}\label{sec:as}

Homology groups and the associated Betti numbers are most easily computed for a class of relatively simple objects called simplicial complexes. A simplicial complex is a structure built from points, lines, triangles, and higher dimensional analogues called \emph{simplices}\footnote{Technically, a \emph{$k$-simplex} $\sigma$ is the smallest convex set that contains its $k+1$ affinely independent vertices. A \emph{face} of $\sigma$ is any simplex spanned by a subset of its vertices. A \emph{simplicial complex} $\mathcal{K}$ is any set of simplices such that if $\sigma\in\mathcal{K}$ is a simplex, then the faces of $\sigma$ also belong to $\mathcal{K}$ and such that any two simplices in $\mathcal{K}$ are either disjoint or intersect in a common face.}. An illustration of a simplicial complex is shown in figure \ref{fig:homology_example}. Of particular interest is the idea of a \emph{filtration} of a simplicial complex $\mathcal{K}$. This is a nested sequence of simplicial complexes $\varnothing\subseteq\mathcal{K}^0\subseteq\mathcal{K}^1\subseteq\cdots\subseteq\mathcal{K}^m=\mathcal{K}$. By computing the homology at each step, we can follow how the topology changes as points, lines, and triangles are filled in. There are different ways to translate the complex reionisation topology into a usable filtration. The most straightforward way to accomplish this task is with $\alpha$-shapes \citep{edelsbrunner83,edelsbrunner94}.

Alpha-shapes are families of geometric constructions that capture the shape of a point set $\mathcal{P}$ over a range of scales. In this paper, we take as our point set the collection of bubble centres. The $\alpha$-shape is then constructed as follows. We start with the \emph{Voronoi tesselation} of $\mathcal{P}$ \citep{okabe92,icke87,weygaert94}. This is a partition of $\mathbb{R}^3$ into cells, one for each point $p\in\mathcal{P}$. The Voronoi cell of $p$ consists of those points $x\in\mathbb{R}^3$ that are at least as close to $p$ as to any other point $q\in\mathcal{P}$. The \emph{Delaunay triangulation} $\mathcal{T}$ of $\mathcal{P}$ is the dual graph of the Voronoi tesselation. Two points in $\mathcal{P}$ are connected by an edge in $\mathcal{T}$ if their Voronoi cells intersect. The Delaunay triangulation $\mathcal{T}$ is a simplicial complex. Its simplices are spanned by the sets of $k+1$ points in $\mathcal{P}$ whose circumscribing sphere does not contain any other point in $\mathcal{P}$. See the first two panels in figure \ref{fig:alpha_shape} for an example. By taking suitable subsets of $\mathcal{T}$, we get a filtration.

For any value of $\alpha\geq0$, we define the \emph{$\alpha$-complex} as a particular subset of the Delaunay triangulation. We draw a ball of radius\footnote{Another commonly used convention is that the radius of the ball is $\sqrt{\alpha}$.} $\alpha$ around each point $p\in\mathcal{P}$. Those simplices of $\mathcal{T}$ that are contained within the union of balls belong to the $\alpha$-complex. The $\alpha$-shape is the union of all simplices in the $\alpha$-complex. As $\alpha$ grows larger, the $\alpha$-shape gets filled in. This is what we see in the last two panels of figure \ref{fig:alpha_shape}. The $\alpha$-shape only changes at discrete values of $\alpha$. By increasing $\alpha$ until the entire Delaunay triangulation is filled in, we produce our desired filtration.

A crucial point is that the bubble network, which we now understand as the union of all closed balls of radius $\alpha$ centred on a point in $\mathcal{P}$, is \emph{homotopy equivalent} to the corresponding $\alpha$-shape. This condition is slightly weaker than being \emph{homeomorphic}, in which case all topological properties of the two shapes would be identical, but it does mean that the shapes can be continuously deformed into each other. In particular, it implies that the bubble network and the $\alpha$-shape have the same number of holes, validating our approach.

\subsection{Weighted $\alpha$-shapes}\label{sec:was}

To model ionisation bubbles of different sizes or born at different times, we need to go beyond the simple $\alpha$-shapes of the previous section. In this case, \emph{weighted $\alpha$-shapes} provide the appropriate filtration set \citep{edelsbrunner92,edelsbrunner95}. Refer to figure \ref{fig:log_normal_slices} for examples of non-uniform bubble networks that require weighted $\alpha$-shapes. This is a generalisation of the above construction, where each point $p$ is assigned a weight $w_p$. We picture a weighted point $(p,w_p)$ as a sphere centred on $p$ with radius $w_p$. Consider the weighted point set $\mathcal{P}$. Let $\mathcal{B}$ be the set of closed balls with boundary in $\mathcal{P}$. The union $\mathcal{F}=\bigcup\mathcal{B}$ of these balls is what we understand as a bubble network.

Define the \emph{weighted distance} from $(p,w_p)$ to $(q,w_q)$ as
\begin{align}
\pi(p,q) = |\!|p-q|\!|^2 - w_p^2 - w_q^2, \label{eq:distfunc}
\end{align}

\noindent
where $|\!|p-q|\!|$ is the Euclidean distance. The \emph{weighted Voronoi cell} of $(p,w_p)$ consists of all unweighted points $x\in\mathbb{R}^3$ whose weighted distance to $p$ is no more than the weighted distance to any other $q\in\mathcal{P}$. The \emph{weighted Delaunay triangulation} is then the dual graph of the \emph{weighted Voronoi tesselation}.

Denote by $\mathcal{F}_\alpha$ the bubble network that is obtained by inflating every sphere $(p,w_p)$ to a sphere $(p,r)$ with radius
\begin{align}
r = \sqrt{\text{sign}(w_p)w_p^2 + \text{sign}(\alpha)\alpha^2}. \label{eq:radius_func}
\end{align}

\noindent
We explicitly allow for negative values of $\alpha$, so that we can both inflate ($\alpha>0$) and deflate ($\alpha<0$) the bubbles. The interpretation of negative weights ($w_p<0$) is explained later. Points with $r^2<0$ are called \emph{redundant} and are omitted. The reason for using the non-linear radius function \eqref{eq:radius_func} is that the resulting Voronoi cells are unchanged when $\alpha$ is varied. It follows that the dual Delaunay triangulations are also independent of $\alpha$, allowing us to build a filtration analogous to the unweighted case.

\begin{figure}
	\centering
	\centering\includegraphics[width=.5\linewidth]{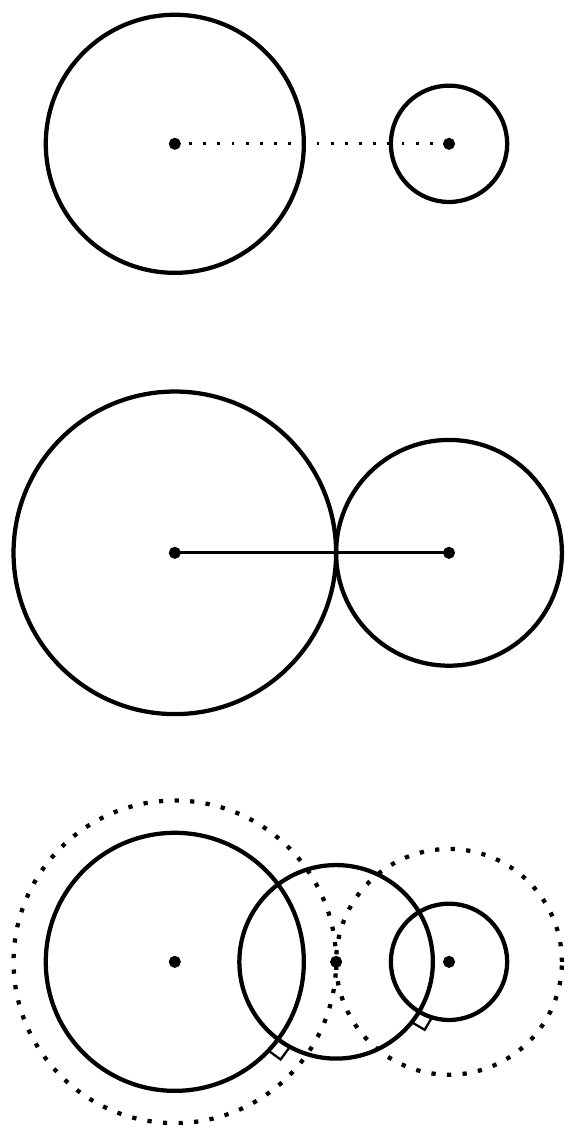}
	\caption{Two weighted points and the simplex spanned by their centres (top). The spheres are $\alpha$-inflated via equation \eqref{eq:radius_func} until they intersect, at which point the edge enters the $\alpha$-complex (middle). If we place a sphere of radius $\alpha$ at the point of intersection, then it is orthogonal to the original spheres (bottom). The size of the edge is $\alpha$.}
	\label{fig:intersection}
\end{figure}

The weighted $\alpha$-complex is constructed as follows. Let $\sigma\in\mathcal{T}$ be a simplex in the weighted Delaunay triangulation. The simplex is part of the $\alpha$-complex if it is the face of another simplex in the complex or if it is ``smaller than $\alpha$''. This agrees with our intuition for the unweighted case, where a simplex was added as soon as the balls were large enough to contain it, but the formal definition requires some thought. We define the \emph{size} $y_\sigma$ of $\sigma$ to be the smallest value of $\alpha$ for which the $\alpha$-inflated spheres centred on its $k+1$ vertices intersect in a point $x$. Equivalently, $y_\sigma$ is the radius of the smallest sphere $x$, such that $\pi(x,p)=0$ for all vertices $p$ of $\sigma$. It may be useful to note that $\pi(x,p)=0$ if and only if the spheres $x$ and $p$ are orthogonal. Now we say that the simplex is part of the $\alpha$-complex if $\alpha\geq y_\sigma$, provided there are no conflicts with other points in $\mathcal{P}$. A conflict occurs if $\pi(x,q)<0$ for any point $q\in\mathcal{P}$ that is not a vertex of $\sigma$. See figure \ref{fig:intersection} for an example where $\sigma$ is a line segment.

We return to the point on negative weights. If we have bubbles at locations $\{p_1,p_2,\dots\}$ born at times $\{\tau_1,\tau_2,\dots\}$, we define the weights by $w_i=-\tau_i$. This ensures that for $\alpha<\tau_i$, the point $p_i$ is redundant, but for $\alpha\geq\tau_i$, we get an inflating bubble with radius $\sqrt{\alpha^2-\tau_i^2}$. A point with negative weight $w_p<0$ is thus interpreted as a bubble born at ``time'' $\alpha=-w_p>0$.

\subsection{Field filtrations}\label{sec:field_filtrations}

Instead of using $\alpha$-shapes, we may wish to build a filtration that directly reflects the properties of some scalar field $f\colon\mathbb{R}^3\to\mathbb{R}$, such as the ionisation fraction field. One way to do this is as follows. We first specify a point set $\mathcal{P}$ at which the field is sampled, such that each vertex $p\in\mathcal{P}$ has an associated field value $f(p)$. As in the $\alpha$-shape method, the filtration consists of subsets of a Delaunay triangulation of $\mathcal{P}$. At any value of $\alpha\in\mathbb{R}$, those vertices $p$ with $f(p)\geq\alpha$, as well as any simplices connecting them, are part of the simplicial complex $\mathcal{K}_\alpha$. Assuming that $f$ is smooth, $\mathcal{K}_\alpha$ only changes when $\alpha$ passes through a critical value of the field $f$. We thus obtain our desired filtration $\varnothing\subseteq\mathcal{K}^0\subseteq\mathcal{K}^1\subseteq\cdots\subseteq\mathcal{K}^m=\mathcal{K}$.

An important question concerns how to choose the point set $\mathcal{P}$ in a way that preserves the topology of the field. When we have discrete samples or measurements of the field, this can be done with the Delaunay Tessellation Field Estimator \texttt{DTFE} \citep{schaap00,weygaert08b,cautun11}. This method has previously been applied to the cosmic density field by \citet{pranav16}. We refer to their paper for more details on this approach.

\subsection{Homology}\label{sec:homology}

Betti numbers are derived from the field of algebraic topology. Algebraic topology is about finding ways of mapping topological spaces to algebraic objects, such as groups. One example, and the one in which we are interested, is that of homology groups. As mentioned before, the idea behind homology is that we can characterise the topology of an object in terms of the cycles or loops that we can draw on it. Equivalently, homology tells us about the boundaries of and holes in a space. Two loops are equivalent when they can be continuously deformed into each other. On the sphere any loop can be contracted to a point, but on the torus there are two classes of non-contractible loops that cannot be deformed into each other. This corresponds to the fact that there are no 1-dimensional holes in the sphere and two distinct 1-dimensional holes in the torus: the hole through the middle and any cross section of the tunnel that runs along the torus.

We can generalise the idea of loops and holes to arbitrary dimensions (see figure \ref{fig:holes}). As previously explained, we have points and gaps ($k=0$), loops and tunnels ($k=1$), and shells and voids ($k=2$). In arbitrary dimensions, we talk about $k$-cycles surrounding $k$-dimensional holes. The homology classes in dimension $k$ can be arranged into a group called the $k$th homology group $\mathcal{H}_k$. The $k$th Betti number $\beta_k$ is the rank of this group. We arrive again at the notion that the $k$th Betti number describes the number of $k$-dimensional holes. In section \ref{sec:hom_example}, we give a little more insight in how these notions are defined for simplicial complexes. Refer to \citet{munkres84,hatcher01} for a textbook introduction.

\subsection{Simplicial homology}\label{sec:hom_example}

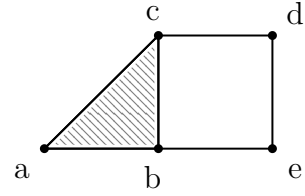
\begin{figure}
	\centering
	\begin{tikzpicture}[thick,scale=1.5]
		\node (a) at (0, 0) {};
		\node (b) at (1, 0) {};
		\node (c) at (1, 1) {};
		\node (d) at (2, 1) {};
		\node (e) at (2, 0) {};
		
		\node at (-.2, -.2) {\Large a};
		\node at (.95, -.25) {\Large b};
		\node at (.95, 1.2) {\Large c};
		\node at (2.2, 1.2) {\Large d};
		\node at (2.2, -.2) {\Large e};
		
		\fill[pattern=north west lines, pattern color=gray] (a.center)--(b.center)--(c.center);		
		\draw[color=white, line width=3pt] (a.center)--(b.center)--(c.center)--(a.center);
		\draw[line width=0.75pt] (a.center)--(b.center)--(c.center)--(a.center);
		\draw[line width=0.75pt] (a.center)--(b.center)--(c.center)--(a.center);
		\draw[line width=0.75pt] (b.center)--(e.center)--(d.center)--(c.center);
		
		\draw[color=black, fill=black] (a) circle (.03);
		\draw[color=black, fill=black] (b) circle (.03);
		\draw[color=black, fill=black] (c) circle (.03);
		\draw[color=black, fill=black] (d) circle (.03);
		\draw[color=black, fill=black] (e) circle (.03);
	\end{tikzpicture}
	\caption{A simplicial complex consisting of five points, six line segments, and one (filled in) triangle.}
	\label{fig:homology_example}
\end{figure}

The homology of a simplicial complex can be defined in terms of chains of simplices. To demonstrate this, consider the simplicial complex in figure \ref{fig:homology_example}. There are five 0-simplices, namely the points $a,\dots,e$. There are six 1-simplices or line segments, which we write as $[a,b]$. There is one 2-simplex, namely the triangle $[a,b,c]$. We start with the observation that these simplices can be chained together. For instance, we could write the path around the triangle as $\sigma=[a,b]+[b,c]+[c,a]$. In general, we call any linear combination of $k$-simplices with integer coefficients (modulo $p$) a $k$-chain. With the operation of addition, the $k$-chains form a free Abelian group\footnote{Every $k$-chain in $\mathcal{C}_k$ is a formal sum of elements of a basis $B$, consisting of the $k$-simplices in the complex. We say that the group is free over $B$.} called the $k$th chain group $\mathcal{C}_k$.

Given a $k$-chain $\sigma$, we can construct a $(k-1)$-chain $\partial\sigma$ called its \emph{boundary}. For instance, the boundary of a line segment is the difference of its endpoints and the boundary of a triangle is the path around it. A chain whose boundary is zero is called a \emph{cycle}. The boundary of the 2-chain $[a,b,c]$ is the 1-chain $\sigma=[a,b]+[b,c]+[c,a]$. The boundary of $\sigma$ is 0, since its endpoints coincide. Hence, $\sigma$ is also a 1-cycle. The path $\tau$ around the square is similarly a $1$-chain and a $1$-cycle, but not a boundary since it does not enclose any triangles. The boundaries and cycles form subgroups of the chain group, denoted as $\mathcal{B}_k$ and $\mathcal{Z}_k$ respectively. 

Two cycles are \emph{homologous} if they differ by a boundary. Visually, this means they surround the same holes. For example, the cycle $\sigma+\tau$ that encircles the combined triangle and square figure is homologous to the cycle $\tau$ that just goes round the square, because the difference $\sigma$ is a boundary. Being homologous is an equivalence relation. All $k$-cycles can thus be partitioned into homology classes. These homology classes form a group called the $k$th homology group $\mathcal{H}_k$. This can also be understood as the factor group $\mathcal{H}_k=\mathcal{Z}_k/\mathcal{B}_k$. Recall that the $k$th Betti number is the rank of $\mathcal{H}_k$. In the example above, there are two independent $1$-cycles namely the path $\sigma$ around the triangle and the path $\tau$ around the square. Thus the 1-cycle group $\mathcal{Z}_1$ has a basis $\{\sigma,\tau\}$ and rank 2. There is only one independent 1-boundary, namely $\sigma$, so the 1-boundary group $\mathcal{B}_1$ has rank 1. Hence, we find that $\beta_1=\text{rank}\;\mathcal{H}_1=\text{rank}\;\mathcal{Z}_1 - \text{rank}\;\mathcal{B}_1=1$. Intuitively, this agrees with the fact that there is one 1-dimensional hole, namely the one enclosed by the square.

\subsection{Persistence diagrams}\label{sec:algo}

As the previous example shows, the problem of identifying the $k$-dimensional holes in a simplicial complex can be solved by finding suitable bases for the cycle and boundary groups $\mathcal{Z}_k$ and $\mathcal{B}_k$. Computationally, it is convenient to do this by representing the boundary operator as a matrix. As an example, consider a simplicial complex consisting of a single triangle $[a,b,c]$ with boundary $\sigma=[a,b]+[b,c]+[c,a]$. Recall that the boundary of an edge is the difference of its endpoints: $\partial[a,b]=b-a$. With respect to the basis $\{a,b,c\}$, we write the boundaries of the 1-simplices as
\begin{align*}
\left[\begin{array}{c|ccc}
&[a,b]&[b,c]&[c,a]\\
\hline
a&-1&0&1\\
b&1&-1&0\\
c&0&1&-1\end{array}\right] \sim \left[\begin{array}{c|ccc}
&[a,b]&[b,c]&\sigma\\
\hline
b-a&1&0&0\\
c-b&0&1&0\\
0&0&0&0\end{array}\right].
\end{align*}

\noindent
On the right, we have brought the matrix to Smith normal form by means of elementary row and column operations (switching rows or columns, multiplying them by a non-zero scalar, and adding multiples of one row or column to another). In this form, we see that $\{b-a,c-b\}$ is a basis for the boundary group $\mathcal{B}_0$ and $\{\sigma\}$ is a basis for the cycle group $\mathcal{Z}_1$.

We use similar techniques to compute the persistent homology of a filtration $\varnothing\subseteq\mathcal{K}^0\subseteq\mathcal{K}^1\subseteq\cdots\subseteq\mathcal{K}^m=\mathcal{K}$. The goal is to identify every hole that appears in the filtration. Each hole first appears as a cycle in some complex $\mathcal{K}^i$. If the hole is still present in $\mathcal{K}$, we assign it a pair $(i,\infty)$. Other holes disappear when they are filled up. This occurs when the corresponding cycle becomes a boundary, say in $\mathcal{K}^j\supseteq\mathcal{K}^i$. In that case, we assign the hole a pair $(i,j)$.

To compute these $(\text{birth}, \text{death})$-pairs, we use the algorithm of \citet{zomorodian05}. The basic idea is as follows. We loop through every simplex $\sigma^j$ in the order in which they appear in the filtration. We maintain matrices akin to the ones above, but suitably generalised to represent the homology of an entire filtration. The matrices are updated incrementally using elementary column operations. For each simplex $\sigma^j$, we first compute its boundary. We then check if the boundary corresponds to a zero column in the boundary matrix. If not, we find the simplex $\sigma^i$ in the boundary that appears latest in the filtration. The set-up then guarantees the existence of a cycle that is born when $\sigma^i$ enters the filtration and becomes a boundary when $\sigma^j$ is added. The corresponding hole is assigned the pair $(i,j)$. A more extensive discussion of a very similar computational paradigm can be found in \citet{pranav16}.

We have implemented this algorithm for $\alpha$-shape filtrations\footnote{Our software is available at \url{http://willemelbers.com/persistent-homology/}.}. Let us make a few practical remarks. The algorithm computes the persistent homology over a finite field $F_p$. This is why we mentioned above that the coefficients of the chains are integers modulo a prime number $p$. We used $p=2$ for the results in this paper. The algorithm also works for larger $p$, but the differences are negligible for our purposes. Secondly, the algorithm outputs a pair $(i,j)$ for each feature, corresponding to the indices of the complexes in which the feature first appears and disappears, respectively. Since each complex $\mathcal{K}^i$ has an associated scale $\alpha^i\in\mathbb{R}$, this is equivalent to computing the $(\alpha_\text{birth},\alpha_\text{death})$ values. Finally, we note that the algorithm works with an efficient data structure. This means we do not actually maintain the boundary matrices, which would be impractical for our application.

\begin{figure*}
	\includegraphics[width=\linewidth]{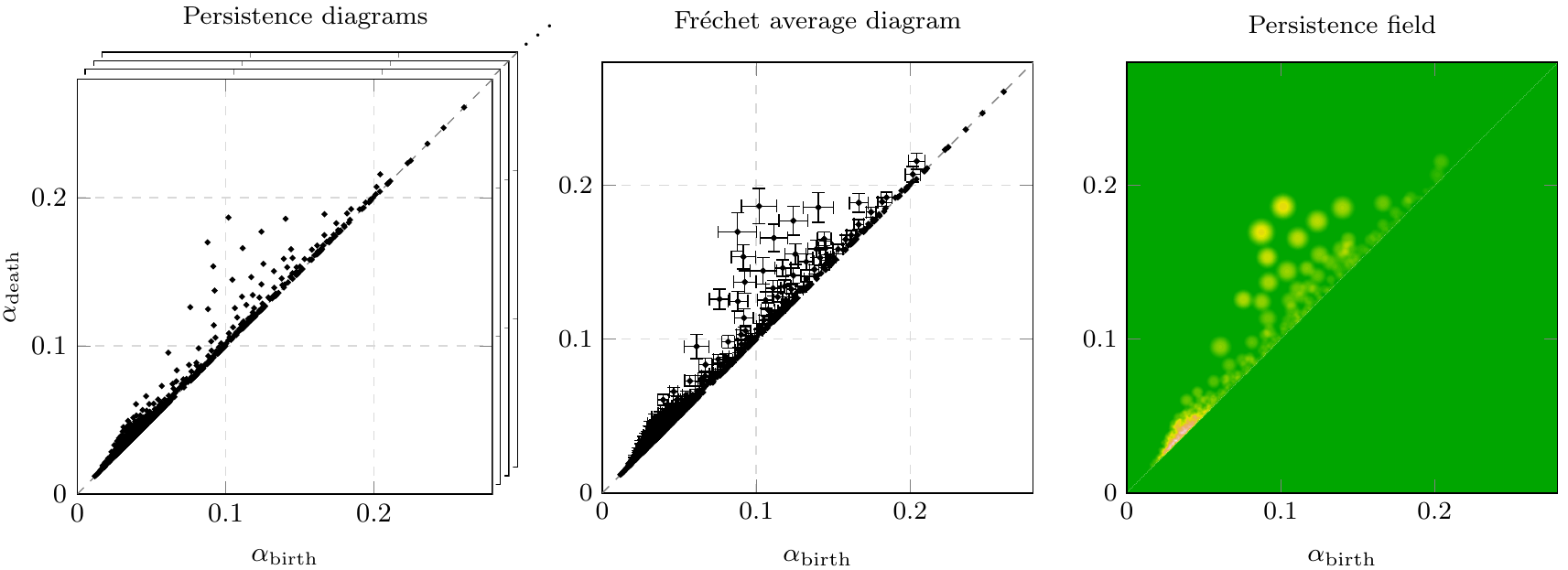}
	\caption{Our pipeline for creating persistence fields. Starting with $n$ realisations of a stochastic process, we obtain a sample of $n$ persistence diagrams $\{X_i\}$. We compute a Fr\'echet average diagram $Y$ and associated variances $\sigma_y^2$ of the points $y\in Y$. These are then used to produce a persistence field.}
	\label{fig:pers_diag}
\end{figure*}

Having successfully computed the $(\text{birth}, \text{death})$-pairs, we can plot the topological features in $(\alpha_\text{birth},\alpha_\text{death})$-space, producing a persistence diagram. See figure \ref{fig:pers_diag} for an example. The horizontal (or vertical) distance of a point to the diagonal is its persistence. This particular example shows the births and deaths of tunnels in a clustered model. The persistence diagram reflects the topology of the model. We see for instance that there are two generations of features: a large number of low-persistence features on small scales and a small number of high-persistence features on large scales. The former correspond to mergers within clusters and the latter to mergers of clusters. See section \ref{sec:clustering} for a detailed discussion of this model.

\subsection{Statistics of persistence diagrams}\label{sec:stats}

If the preceding theory is to be applied to real world data, we must be able to handle experimental uncertainties. Even when dealing with simulations, a statistical approach is highly preferable. In this paper, the set-up is as follows. For each of the phenomenological models treated in section \ref{sec:results}, we generate $n$ random bubble networks and compute one persistence diagram $X_i$ for each realisation $i=1,\dots,n$. We are looking for appropriate summary statistics of the sample $S=\{X_i\}$.

To describe the homology from a statistical point of view, we therefore consider the space $\mathcal{D}$ of persistence diagrams. A persistence diagram is nothing more than a collection of $(\alpha_\text{birth},\alpha_\text{death})$-pairs, but we need an additional technical condition to ensure that $\mathcal{D}$ is a well-behaved probability space. Formally then, we define a \emph{persistence diagram} as a countable set of finitely many points $x\in\mathbb{R}^2$ together with infinitely many copies of the diagonal $\Delta=\{(x,y)\in\mathbb{R}^2\mid x=y\}$. In that case, $\mathcal{D}$ is a complete and separable metric space on which probability measures, expectation values, and variances can be defined \citep{mileyko11}. In this paper, we use the $L^2$-Wasserstein metric \citep{turner14}:
\begin{align}
d(X,Y) = \left[\text{inf}_{\phi\colon X\to Y} \sum_{x\in X} |\!| x-\phi(x)|\!|^2\right]^{1/2}.
\end{align}

\noindent
To compute the distance between two persistence diagrams $X,Y\in\mathcal{D}$, we need to consider all bijections $\phi\colon X\to Y$. These are one-to-one maps that match each point $x\in X$ with a point $y\in Y$ and vice versa. Here, we treat the diagonal $\Delta$ as a point that can be matched either with an off-diagonal point or with another copy of the diagonal. Given such a matching $\phi$, the distance $|\!| x-\phi(x)|\!|$ is simply the Euclidean distance from $x$ to its partner $\phi(x)$. We specify that the distance $x$-$\Delta$ is the distance from $x$ to the closest point on the diagonal and that the distance $\Delta$-$\Delta$ is zero. We refer to a bijection $\phi$ that minimises the total squared distance as an \emph{optimal matching} between $X$ and $Y$. Finding such a matching is a form of the assignment problem, which can be solved with the Hungarian algorithm or the auction algorithm of Bertsekas \citep{kerber17}. The $L^2$-Wasserstein distance $d(X,Y)$ is then the square root of the minimum total squared distance.

Given some probability measure $\rho\subset\mathcal{D}$, we define the Fr\'echet function
\begin{align}
F\colon \mathcal{D}\to\mathbb{R}, \;\,\;\,\;\,\;\,\;\,\;\, F(Y) = \int_{\mathcal{D}}d(X,Y)^2\mathrm{d}\rho(X).
\end{align}

\noindent
In the case of a finite sample $S=\{X_i\}\subset\mathcal{D}$, we have $\rho(X)=n^{-1}\delta_S(X)$ and this becomes
\begin{align}
F(Y) = \frac{1}{n}\sum_{i=1}^n d(Y,X_i)^2.
\end{align}

\noindent
A \emph{Fr\'echet mean} of the sample $S$ is a diagram $Y$ that minimises $F(Y)$. In general, this is not unique because $F$ can have multiple minimisers. The \emph{Fr\' echet variance} of $S$ is $F(Y)$. This is a measure of the uncertainty in the sample. If we let $\phi_i\colon Y\to X_i$ be an optimal matching of $Y$ with $X_i$, we can write this as
\begin{align}
F(Y) = \frac{1}{n}\sum_{i=1}^n d(Y,X_i)^2 = \frac{1}{n}\sum_{i=1}^n\sum_{y\in Y} |\!| y-\phi_i(y)|\!|^2.
\end{align}

\noindent
We can thus attribute a part
\begin{align}
\sigma_y^2 = \frac{1}{n}\sum_{i=1}^n |\!| y-\phi_i(y)|\!|^2
\end{align}

\noindent
of the uncertainty to each point $y\in Y$. Unlike the total Fr\'echet variance, this attribution is again not unique, because there can be multiple optimal matchings. However, the generic case is that the assignment problem does have a unique optimal solution, so we ignore this possibility here.

Given a sample of diagrams, a local minimum of $F$ can be found in finite time \citep{turner14}. The mean and variance of a sample can be combined into a \emph{persistence field}, which we discuss further below.

\subsection{Persistence fields}\label{sec:images}

In our analysis, we display the statistics of a sample of persistence diagrams $\{X_i\}$ with a \emph{persistence field}, based on a similar but distinct representation proposed by \citet{adams17}. The goal is to create a visualisation of the persistence data that satisfies a number of objectives. The image should
\begin{enumerate}[label=(\roman*), wide=0pt, widest=99,leftmargin=\parindent, labelsep=*]
	\item resemble the underlying persistence diagrams,
	\item reflect the uncertainty in the sample,
	\item be stable with respect to noise in the data,
	\item reflect the number of topological features,
	\item show both rare high-persistence features and common low-persistence features.
\end{enumerate}

The first two goals suggest that we use a Fr\'echet average $Y$ of the sample diagrams (see section \ref{sec:stats}). This also gives us a measure of the uncertainty $\sigma_y^2$ of each feature $y\in Y$. The third and fourth goals suggest some kind of kernel density estimation or smoothing of the average diagram. This creates a difficulty, because those features that are most significant are also extremely rare and are washed out in any kernel density estimate. We therefore assign each feature $y\in Y$ a weight $w_y$ proportional to the square root of its persistence. We then smooth $Y$ with a Tri-cube kernel
\begin{align}
K(r) = \left(1-r^3\right)^3, \;\,\;\,\;\,\;\, 0\leq r\leq 1.
\end{align}

\noindent
The persistence field $f\colon\mathbb{R}^2\to\mathbb{R}$ is then
\begin{align}
f(x) = \sum_{y\in Y} w_y K\big(|\!|x-y|\!|/(b\sigma_y)\big),
\end{align}

\noindent
where $b$ is the bandwidth. The Tri-cube kernel has a relatively flat top so that clearly distinguishable features resemble a disk with radius $\sim b\sigma_y$. Our pipeline is illustrated in figure \ref{fig:pers_diag}.

\begin{figure*}
  \centering
  \subfloat[$\alpha=-0.05$]{
    \centering\includegraphics[width=.325\linewidth]{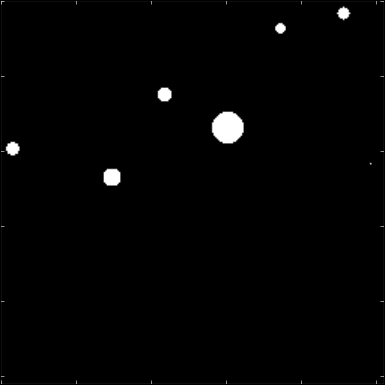}
	\label{fig:small_log_normal_alphamin}
  }
  \subfloat[$\alpha=0$]{
    \centering\includegraphics[width=.325\linewidth]{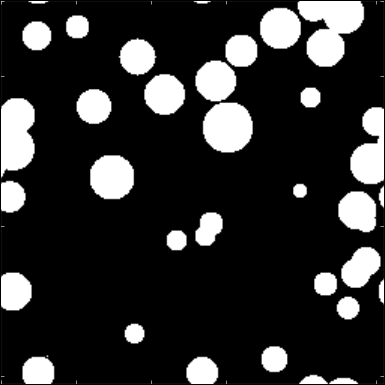}
	\label{fig:small_log_normal_alphazero}
  }
  \subfloat[$\alpha=0.05$]{
    \centering\includegraphics[width=.325\linewidth]{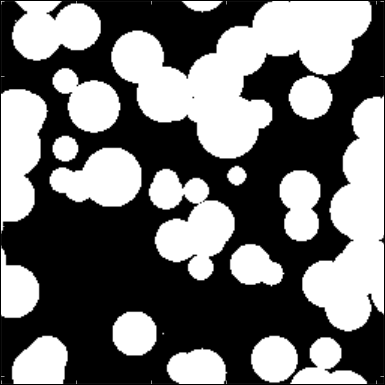}
	\label{fig:small_log_normal_alphaplus}
  }
  \caption{Slices of a non-uniform bubble network with log-normal bubble sizes $(\mu=-3.0,\sigma=0.10)$, for different values of $\alpha$. The median bubble radius is $0.05$. This means that at $\alpha=-0.05$, half of all bubbles are \emph{redundant} and have yet to appear. Those that have appeared are deflated. At $\alpha=0$, all bubbles are present and have exactly their log-normal radius. At $\alpha=0.05$, the bubbles have been inflated.}
  \label{fig:log_normal_alpha_slices}
\end{figure*}

\section{Structural filtrations}\label{sec:filtrations}

We study bottom-up filtrations of the ionisation bubble network. By considering filtrations along different dimensions (e.g. length scale or time), we can study different aspects of the ionisation topology. The topology is characterised in terms of the births and deaths of topological features at every scale. The precise meaning of this scale, and the interpretation of topological persistence, depends on the dimension along which we build our filtration. However, common to all filtrations is the interpretation of the features themselves. By topological features we mean islands, tunnels, and voids in the ionisation field. The connected components or islands are simply the ionisation bubbles or the distinct regions of ionised material. The tunnels are neutral filaments that pierce through the ionisation bubble network. Voids are patches of neutral hydrogen enclosed by ionised material. Note that these are voids in the ionisation field, which often correspond to overdensities. They are distinct from voids in the cosmic density field, which correspond to underdensities.

\subsection{Spatial structure}

First, we consider a snapshot of the ionisation field at a fixed redshift. As input, we need the locations $\{x_1,x_2,\dots\}$ of the ionising sources. We also need to specify the radius $r_i$ of the ionised region surrounding the source at $x_i$. These data could be the output of a semi-numerical model or obtained by applying granulometry \citep{kakiichi17} to the ionisation map of a full radiative transfer simulation, or to 21-cm tomographic images (see section \ref{sec:sources}). The bubble size distribution could also be constrained by observation through other means \citep{friedrich11,lin16,giri17}.

Associate a weight $w_i=r_i$ with the source at $x_i$. We then use the weighted point set $\mathcal{P}=\{(x_1,w_1),(x_2,w_2),\dots\}$ as the basis for a weighted $\alpha$-complex. The filtration consists of the (finitely many) distinct $\alpha$-shapes obtained as we increase the scale from $\alpha=-\infty$ to $\alpha=\infty$. With this filtration, we probe the connectivity structure of the ionisation field at a particular redshift. This is particularly useful for investigating the multi-scale nature of the largest ionised region that arises as a result of the hierarchical build-up of structure. The persistence of a feature has its usual interpretation as topological significance\footnote{When $\alpha$-shapes are used in pattern recognition, persistence is useful as a criterion for filtering out noise \citep{edelsbrunner10}.}.

This is the only type of filtration that involves both negative and positive values of $\alpha$. It is worthwhile to pause here and understand why. At $\alpha=0$, the bubbles have precisely their prescribed radius $\sqrt{r_i^2+\alpha^2}=r_i$. Negative values of $\alpha$ correspond to deflating the bubbles. A bubble disappears when its deflated radius becomes zero, which happens at $\alpha=-r_i$. Therefore, the bubble size distribution is encoded in the persistent homology of the spatial filtration for negative values of $\alpha$. Positive values of $\alpha$ correspond to inflating the bubbles. Among other things, this allows us to determine the topological significance of features that exist in the bubble network at $\alpha=0$, by considering at what scale $\alpha_\text{death}$ the feature disappears. In figure \ref{fig:log_normal_alpha_slices}, we see the same bubble network at negative, zero, and positive values of $\alpha$.

Analysing the spatial filtration also reveals rich information about the connectivity structure of the bubble network. In particular, the fractal nature of the ionised region should be apparent at both positive and negative scales. Small bubbles that have been absorbed into larger bubbles at $\alpha=0$ must have merged at some $\alpha<0$. Similarly, clusters that are separated at $\alpha=0$ will merge when the bubbles are sufficiently inflated, affecting the homology at $\alpha>0$.

\subsection{Bubble dynamics}\label{sec:bubble_dynamics}

A second interpretation of the above filtration is obtained by taking cosmic time $t$ as our filtration parameter: $\alpha=t$. Again, we require the source locations $\{x_1,x_2,\dots\}$. Let $\tau_i$ be the formation time of the bubble at $x_i$ and define its weight through $w_i=-\tau_i$. We then consider the weighted $\alpha$-complex with point set $\mathcal{P}=\{(x_1,w_1),(x_2,w_2),\dots\}$. The filtration consists of the distinct $\alpha$-shapes that we find as we increase time from $t=0$ to $t=\infty$. The reason for taking negative weights is that it allows us to start at $t=0$, such that the source at $x_i$ is born when $t=\tau_i$. The bubbles expand when $t$ is increased, simulating the process of reionisation. Because weighted $\alpha$-shapes are based on the distance function \eqref{eq:distfunc}, this technique requires all bubbles to grow at a non-linear rate $\sim\sqrt{t^2-\tau_i^2}$ and assumes that the bubbles are spherical. Despite these limitations, this simple toy model already displays many of the qualitative features of reionisation. A major conceptual advantage of the $\alpha$-shape method is therefore that we can take $\alpha$ as a measure of time, allowing us to display the entire topological history of the ionisation field in one figure. Moreover, the persistence of a feature can be interpreted as its lifetime.\\

To circumvent the limitations of the $\alpha$-shape method, we also propose an alternative method that makes use of field filtrations (section \ref{sec:field_filtrations}). This allows us to consider two further filtrations.

\subsection{Ionisation gradient}

Up until this point, we assumed a binary ionisation field and probed the topology along the dimensions of time and space. A third dimension would be the ionisation fraction itself. We again start with a time slice of the ionisation history, but now build a filtration by taking superlevel sets of the ionisation fraction field. This can be done as follows. First, we need a set of vertices $p\in\mathcal{P}$ at which the ionisation field is probed. We then construct a Delaunay triangulation $\mathcal{T}$ of $\mathcal{P}$ and a linearly interpolated ionisation field with \texttt{DTFE} \citep{cautun11}. Using these data as input, we compute the persistent Betti numbers of the field filtration of $\mathcal{T}$. The methodology is essentially the same as in \citet{pranav16}, except that we replace the matter density field with the ionisation fraction field. This method is useful for studying regions that are in the process of being ionised at a particular moment. The persistence of a feature is now interpreted as the differential ionisation fraction of the hole. For instance, the persistence of an opening tells us about the ionisation state of the enclosed filament.

\subsection{Full evolution}

Given the output of a more realistic model, we can also build a filtration simply by playing back the ionisation history. To do this, we assign every vertex $p\in\mathcal{P}$ a value corresponding to the redshift at which that point was first considered to be part of an ionised region. The filtration is then built by taking superlevel sets of this field. A first goal will be to compare the topology of a full radiative transfer simulation with that of the bubble dynamics model considered in secion \ref{sec:bubble_dynamics}. One caveat that remains is that filtrations are strictly nested sequences, so regions that recombine cannot be handled easily.

\section{Source properties}\label{sec:sources}

In this paper, we study the spatial structure and dynamics of the ionisation bubble network using a number of phenomenological models. In each of our models, $N$ sources are placed in a periodic unit box $X\subset\mathbb{R}^3$. The $\alpha$-shapes are then computed from the set of source locations using the computer package \texttt{CGAL} \citep{cgal16,cgal16b}. The topological properties of the network are computed using the algorithm discussed in section \ref{sec:algo}. We use different methods of generating the bubble locations and weights in order to illustrate different aspects of the reionisation process. To this end, we need to specify some of the properties of the ionising sources.

\subsection{Source distribution}

\begin{figure*}
  \centering
  \subfloat[Clustered Model]{
    \centering\includegraphics[width=.325\linewidth]{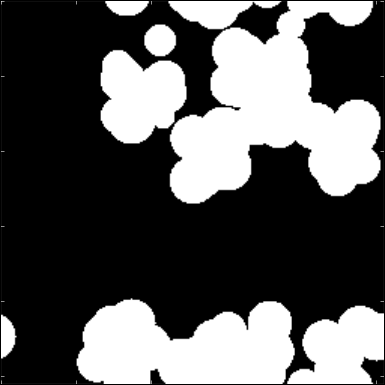}
	\label{fig:correlated}
  }
  \subfloat[Poisson Model]{
    \centering\includegraphics[width=.325\linewidth]{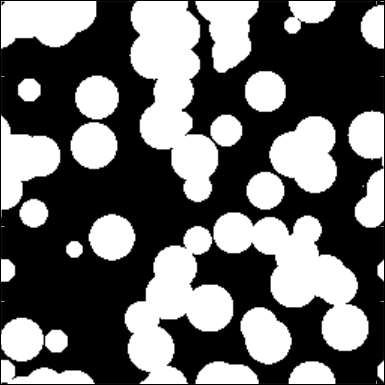}
	\label{fig:poisson}
  }
  \subfloat[Anti-Clustered Model]{
    \centering\includegraphics[width=.325\linewidth]{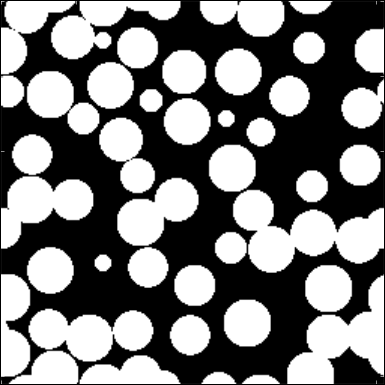}
	\label{fig:anti_correlated}
  }
  \caption{Slices of uniformly sized bubble networks generated with three different point processes. All three pictures correspond to one particular value of $\alpha=0.06$. Because these are uniform bubble networks, all bubbles have the same radius $\alpha$.}
  \label{fig:clustering}
\end{figure*}

The first property that is needed is the spatial distribution of the ionising sources. In realistic models, the source locations will be correlated with the matter density field. Here, we generate the locations with three spatial point processes. In section \ref{sec:clustering}, we use these toy models to demonstrate how the source distribution is reflected in the topology. To isolate the role of the spatial distribution, we assume that all bubbles are born at the same time, in which case the resulting bubble networks are uniformly sized. This means they can be generated with unweighted $\alpha$-shapes. See figure \ref{fig:clustering} for slices through uniform bubble networks generated according to the different point processes discussed below.

\subsubsection{Poisson model}

The simplest way to generate the locations is with a Poisson point process with intensity $\Lambda=N$. The actual number of sources is a Poisson random variable, but we tweak the process to ensure that precisely $N$ sources are generated.

\begin{figure*}
  \centering
  \subfloat[Log-normal Model $(-4.0,0.50)$]{
    \centering\includegraphics[width=.325\linewidth]{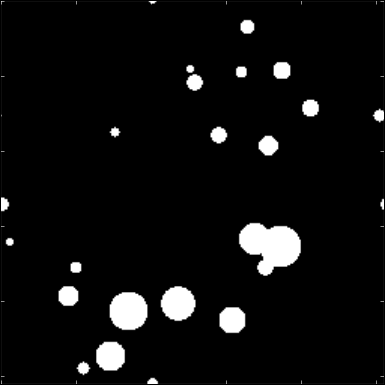}
	\label{fig:small_log_normal}
  }
  \subfloat[Log-normal Model $(-3.0,0.10)$]{
    \centering\includegraphics[width=.325\linewidth]{log_normal_3d/medium.png}
	\label{fig:medium_log_normal}
  }
  \subfloat[Log-normal Model $(-2.5,0.05)$]{
    \centering\includegraphics[width=.325\linewidth]{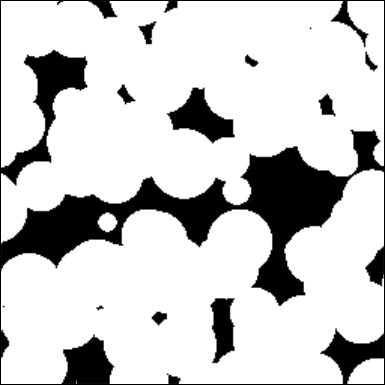}
	\label{fig:large_log_normal}
  }
  \caption{Slices of non-uniform bubble networks generated with log-normal $(\mu,\sigma)$ bubble sizes. All three pictures are taken at $\alpha=0$, so the bubbles are neither inflated nor deflated. Compare panel \ref{fig:medium_log_normal} with figure \ref{fig:log_normal_alpha_slices}, where the same network is depicted for different $\alpha$.}
  \label{fig:log_normal_slices}
\end{figure*}

\subsubsection{Clustered model}

The locations in the Clustered Model are generated with a Neyman-Scott process \citep{neyman58}. The model is described by two parameters $K$ and $\lambda$ in addition to the number of centres $N$. Initially, $K$ cluster centres are generated with a Poisson point process. Subsequently, $N/K$ (rounded to the nearest integer) sources are placed with another Poisson process in a sphere of radius $\lambda K^{-1/3}$ around each of the $K$ initial locations. In this way, $K$ clusters of $N/K$ sources are created.

\subsubsection{Anti-Clustered model}

The Anti-Clustered Model uses a repulsive point process and is described by two parameters: the number of centres $N$ and the minimum centre distance $\lambda$. Sources are generated with a Poisson point process and rejected if they fail the minimum distance requirement until $N$ centres have been produced.\\

\subsection{Bubble size}

When studying the spatial structure of the bubble network, the size distribution of the ionisation bubbles is an important factor. Guided by analytical predictions, many authors have found an approximate log-normal bubble size distribution \citep{furlanetto04,furlanetto05,mcquinn07,mesinger07,friedrich11,zahn11,lin16}. The distribution is expected to peak at a characteristic scale that increases and has a variance that decreases as reionisation progresses and bubbles merge. This motivates the following phenomenological log-normal model \citep{coles91}.

\subsubsection{Log-normal model}

In the log-normal model, the source locations $x_i$ are generated with a Poisson process and the bubble sizes $r_i$ are sampled from a log-normal distribution with parameters $\mu$ and $\sigma$. Figure \ref{fig:log_normal_slices} shows slices through the resulting bubble networks for different values of $(\mu,\sigma)$. This model is used in section \ref{sec:sizes} to investigate how a changing size distribution is reflected in the topology.


\subsubsection{Granulometry}

The $\alpha$-shape method can also be applied to more realistic models of reionisation. Since we need to specify bubble centres $x_i$ and radii $r_i$, we need to find a way to capture the ionised regions in terms of spherical ionisation bubbles. One convenient way to do this is with granulometry \citep{kakiichi17}, which is based on a mathematically well-defined notion of sieving. Applying this technique to tomographic 21-cm images is a promising pathway for the application of our formalism to observation.

\subsection{Bubble age}

When we study bubble dynamics, we also need to specify the bubble formation times $\tau_i$. In realistic models, formation times depend on the matter density field and physical properties of reionisation, such as the local requirements for source formation. In this case, the spatial distribution of the bubbles and their formation times will be related, affecting the topology of the resulting ionisation field.

\subsubsection{Constant Rate Model}

In section \ref{sec:stages}, we study the following model in which the number $N_\text{born}(t)$ of bubbles that have been born at time $t$ increases at a constant rate: $\dot{N}_\text{born}=\text{const}$ until $t=T$, after which the source production turns off. In this model, the bubble locations $x_i$ are chosen with a Poisson process. Hence, the spatial distribution and formation times are indepedent. The formation time $\tau_i$ of the bubble at $x_i$ is sampled from a uniform distribution $U(0,T)$. As we use weighted $\alpha$-shapes to model the bubble networks, we set the radius $r_i(t)$ of the bubble with centre $x_i$ at time $t$ equal to
\begin{align}
r_i(t) = \begin{cases}
0 & \text{if}\;\,\;\, t<\tau_i,\\
\sqrt{t^2-\tau_i^2} & \text{otherwise}.
\end{cases} \label{eq:variable_alpha}
\end{align}

\noindent
This means that the average bubble radius at times $t<T$ will be
\begin{align*}
\langle r(t)\mid\text{alive}\rangle = \int_0^t \frac{\sqrt{t^2-\tau^2}}{t}\mathrm{d}\tau = \frac{\pi t}{4} \approx 0.785t,
\end{align*}

\noindent
whereas the average bubble radius for later times $t\geq T$ is
\begin{align*}
\langle r(t)\rangle &= \frac{1}{2T}\left[Tx+t^2\arctan\left(Tx^{-1}\right)\right] \approx t \;\,\;\,\;\,\text{for }t\gg T,
\end{align*}

\noindent
where $x=\sqrt{t^2-T^2}$. Hence, the average bubble expands at a rate $\dot{r} = 0.785$ initially, after which it approaches $\dot{r}=1$ asymptotically. Different trajectories of $\langle r(t)\rangle$ could be effected by sampling $\tau_i$ from different distributions. However, our methodology means that we have to use the piecewise function \eqref{eq:variable_alpha}.

\section{Results}\label{sec:results}

We now use the models of the preceding sections to demonstrate how different aspects of the reionisation process affect the homology of bubble network filtrations. In section \ref{sec:stages}, we show how the different stages of reionisation can be identified. We then investigate the effect of the spatial distribution of the sources in section \ref{sec:clustering}.  Finally, we consider the effect of the bubble size distribution in section \ref{sec:sizes}.

\subsection{Temporal filtrations}

We start with a number of temporal filtrations. In section \ref{sec:stages}, we study the Constant Rate Model in which the bubbles are born at a constant rate between $t=0$ and $t=T$, after which source production is turned off. In the models considered in section \ref{sec:clustering}, all bubbles are born at $t=0$. As the resulting bubble networks are uniformly sized, these latter models could also be interpreted as spatial filtrations.

\subsubsection{Stages of reionisation}\label{sec:stages}

\begin{table}
	\centering
	\caption{Different epochs in the $N=500$ Constant Rate Model.}
	\begin{tabular}{p{.09\textwidth} p{.2\textwidth} r}
		\hline
		\textbf{Epoch} & \textbf{Ends when} & \textbf{Time}\\
		\hline
		Pre-overlap & 10\% of bubbles overlap & $t=0.038$\\
		Overlap & 97\% of bubbles overlap & $t=0.097$\\
		Filament & Patches outnumber tunnels & $t=0.126$\\
		Patch & Reionisation is complete & $t=0.186$\\
		\hline
		\hline
	\end{tabular}
	\label{tab:epochs}
\end{table}

\begin{figure*}
  \includegraphics[width=\linewidth]{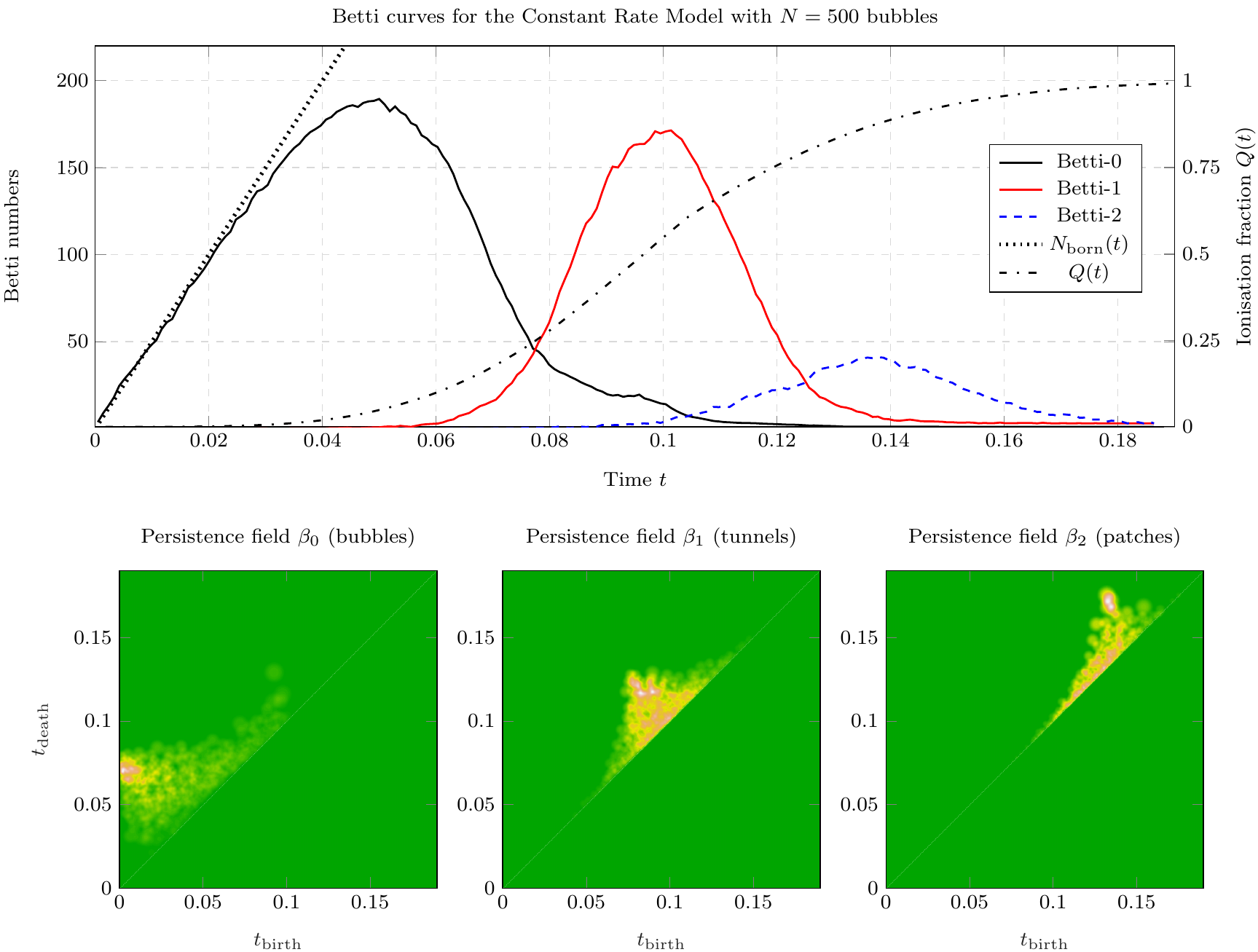}
  \caption{Persistent homology of the Constant Rate Model with $N=500$ bubbles. In the top panel, we see the number of ionisation bubbles ($\beta_0$), neutral filaments ($\beta_1$), and neutral patches ($\beta_2$) alive at any time $t$. Also shown is the number $N_\text{born}(t)$ of bubbles that have been born and the global ionisation fraction $Q(t)$. In the bottom panels, we see the persistence fields showing the births and deaths of all topological features.}
  \label{fig:uniform_results}
\end{figure*}

\begin{figure*}
  \includegraphics[width=\linewidth]{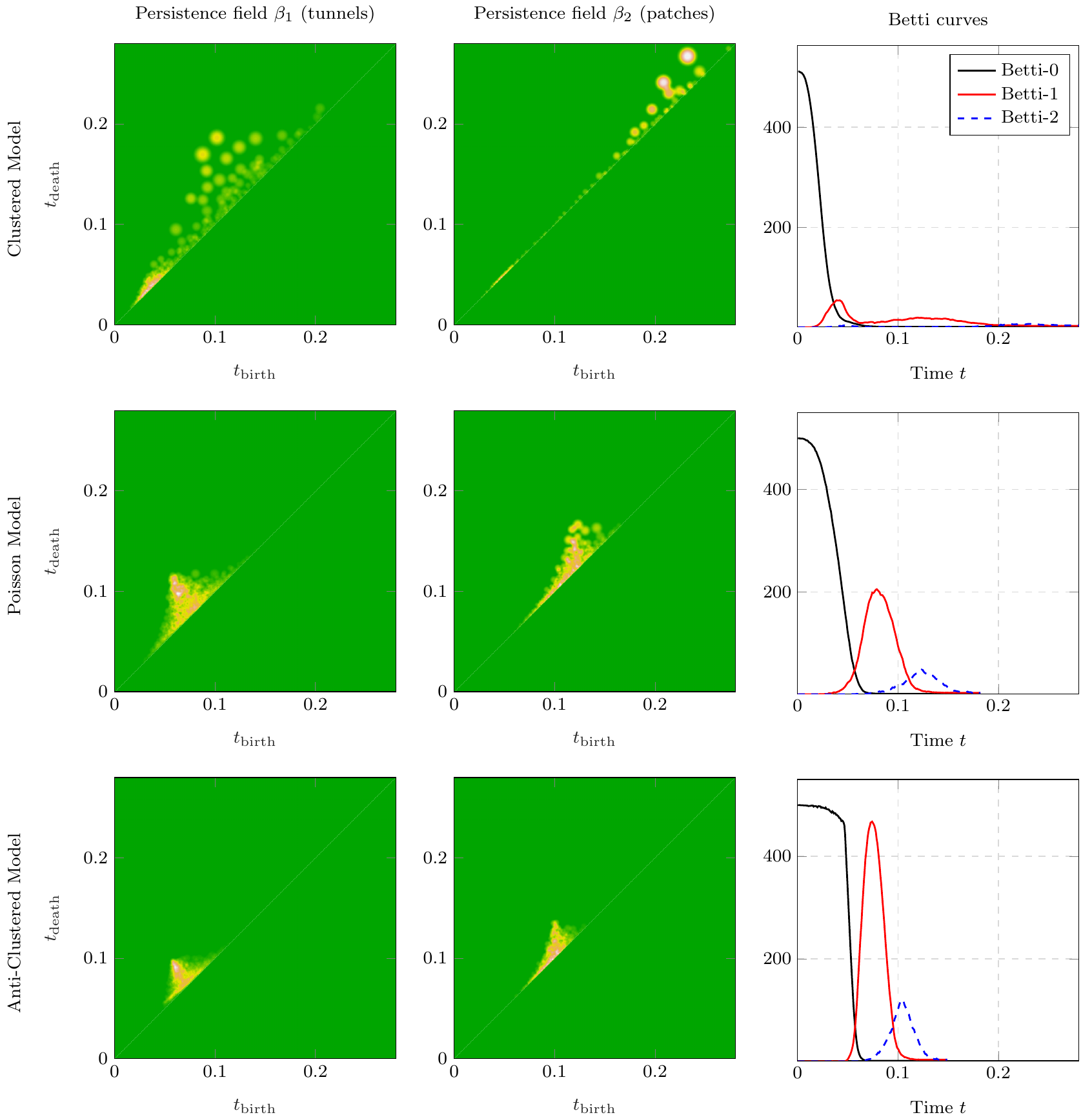}
  \caption{The impact of clustering on persistent homology. The features in the Clustered Model (top) are rare, but far more persistent and spread out compared to the Anti-Clustered Model (bottom). The Poisson Model (middle) is an intermediate case. Notice also that there are two generations of features in the Clustered Model. The early low-persistence features correspond to mergers within clusters and the late high-persistence features correspond to mergers of clusters.}
  \label{fig:clustering_results}
\end{figure*}

\begin{figure*}
  \includegraphics[width=\linewidth]{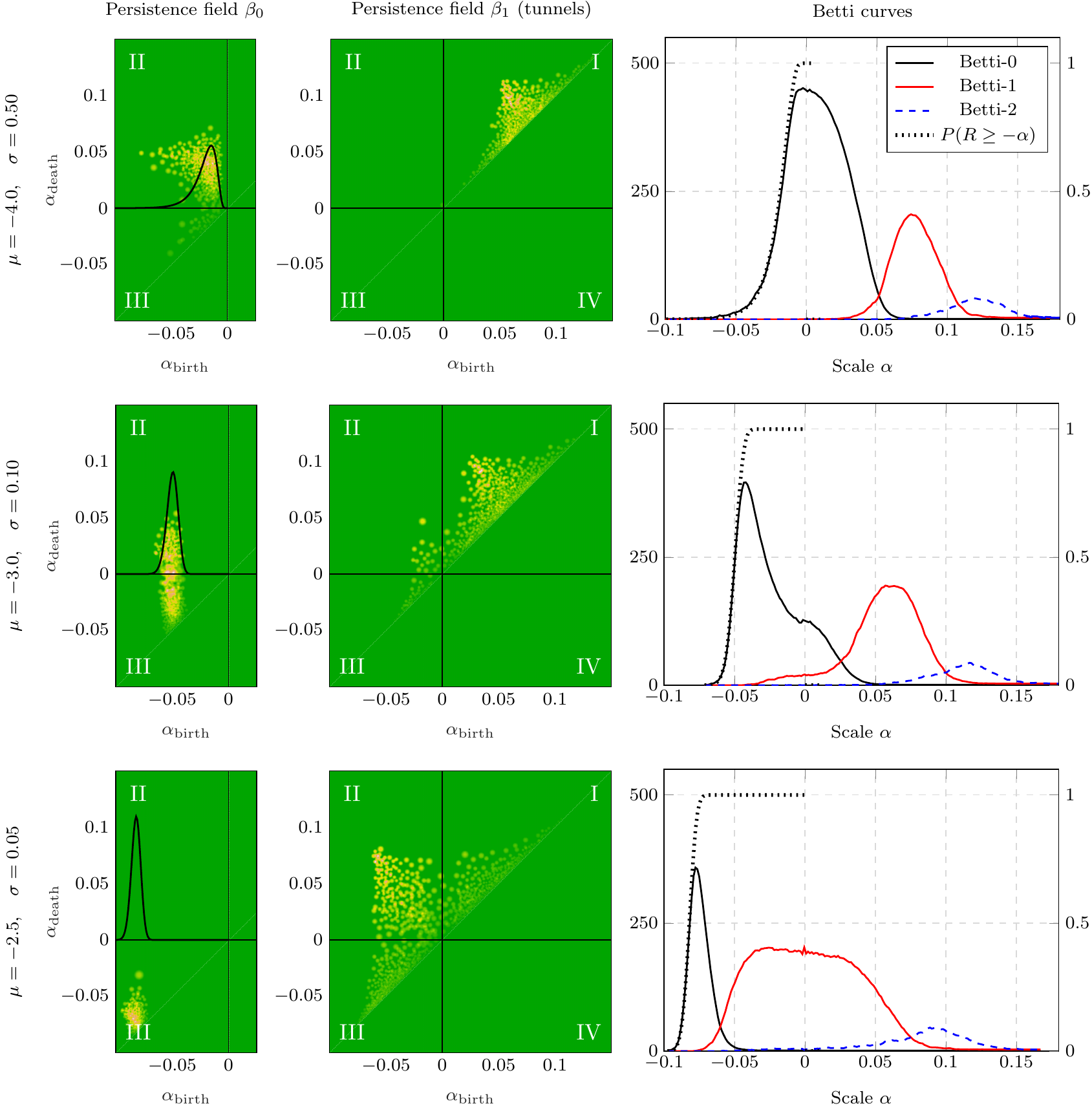}
  \caption{Persistent homology of a bubble network with log-normally distributed bubble sizes, for different values of $\mu$ and $\sigma$. Bubbles are smallest for $\mu=-4.0$ (top) and largest for $\mu=-2.5$ (bottom). The log-normal probability density is overlaid on the $\beta_0$-persistence fields (left). The cumulative distribution function is overlaid on the Betti curves (right). Features in quadrant II of the persistence fields are alive at $\alpha=0$.}
  \label{fig:log_normal_results}
\end{figure*}

We study the different stages of reionisation with a temporal filtration of the Constant Rate Model with $N=500$ bubbles. The bubbles are born at a constant rate over a time interval $[0,T]$ with $T=0.10$. The results, averaged over 10 realisations, are shown in figure \ref{fig:uniform_results}. In the top panel, we see the Betti curves describing the number of ionisation bubbles (Betti-0, solid black), neutral filaments (Betti-1, solid red), and neutral patches (Betti-2, dashed blue). We have overlaid the global ionisation fraction $Q(t)$, which is the fraction of total volume occupied by ionisation bubbles (dot-dashed). The number of ionised regions $\beta_0$ starts at 0 and initially just tracks the number
\begin{align*}
N_\text{born}(t)=\frac{Nt}{T}
\end{align*}

\noindent
of bubbles that have been born (dotted line). After a while, the slope of the $\beta_0$-curve tapers off as bubbles start to overlap and merge. We can therefore use $\beta_0$ as a measure of the degree of overlap. Notice that the ionisation fraction $Q(t)$ only starts to incline appreciably once the number of distinct ionised regions has decreased by 10\% at $t=0.038$. This could be chosen as the end of the pre-overlap stage.

During the overlap stage, the $\beta_0$-curve never reaches far above 200 because any newborn bubbles are immediately fed into larger existing structures. Furthermore, because new bubbles are being born at a constant rate up to $t=0.10$, the $\beta_0$-curve is skewed very much to the right and has a long and fat tail. The ionisation rate reaches a maximum at $t=0.097$. At this point, $\beta_0$ has decreased to just over $3\%$ of its initial value, so this could reasonably be chosen as the end overlap stage. During the subsequent post-overlap stage, the remaining neutral islands are attacked from the outside. Interestingly, most of the higher dimensional structure only appears past this point. First, the number $\beta_1$ of tunnels increases as bubbles begin to overlap that were already connected, forming 1-cycles. These tunnels surround neutral filaments that pierce through the ionisation bubble network. When the filaments are ionised and tunnels begin to be filled up, $\beta_1$ decreases. Meanwhile, $\beta_2$ increases as bubbles start to enclose an increasing number of neutral patches. After $t=0.126$, the patches outnumber the tunnels. Finally, the patches get ionised as well and the Betti numbers reach their final values: $\beta_0=1,\beta_1=3,\beta_2=3$. These values are an artefact of the non-trivial topology of the periodic simulation box $X$, but are negligible compared to the dozens of features found at earlier times.

Looking at the persistence diagrams in the bottom row of figure \ref{fig:uniform_results}, we see that the majority of features are short-lived (close to the diagonal). Nevertheless, a large number of tunnels that are born around $t=0.07$ survive until $t=0.10$, although none live past $t=0.13$. Furthermore, most of the patches that are born before $t=0.13$ die very young, but a large number of patches that are born at $t=0.14$ survive until $t=0.17$. We thus identify two additional topologically significant epochs past the pre-overlap stage and the overlap stage, which may rightly be called the ``filament stage'' and ``patch stage''. These topological characteristics are not apparent from the geometry or ionisation history $Q(t)$. We summarise our criteria for the four stages in table \ref{tab:epochs}.

We also have a persistence field for $\beta_0$, which shows that the longest-living ionised regions emerge at $t=0$, though even at $t=0.10$ some bubbles are born that survive for a relatively long time before being absorbed into larger structures.

\subsubsection{Source distribution}\label{sec:clustering}

To illustrate how the source distribution affects the topology, we study three different models with $N=500$ uniformly sized bubbles placed at random locations: a clustered model, an anti-clustered model, and a Poisson model. A visual inspection of the resulting bubble networks shown in figure \ref{fig:clustering} is quite revealing. We see that the Poisson Model is an intermediate case between two extremes. The bubble network produced by the Clustered Model resembles a two-phase medium consisting of large clusters of bubbles and neutral oceans utterly devoid of bubbles. As a result, the clustered regions are rapidly ionised, but the neutral regions resist ionisation for a long time. At the other extreme, the Anti-Clustered Model produces bubbles appearing in an almost crystal-like pattern. For a long time, these bubbles can freely expand in every direction and when the bubbles finally overlap, the box is almost completely ionised.

To produce these bubble networks, we choose rather extreme model parameters. For the Clustered Model, we generate $K=32$ superclusters with a characteristic size $\lambda K^{-1/3}=0.08$ with $\lambda=0.25$. For the Anti-Clustered Model, we place the sources at least a distance $\lambda=0.094$ from any other source. The results averaged over 10 realisations are shown in figure \ref{fig:clustering_results}. The differences are conspicuous. First, consider the Betti curves in the third column. In contrast to the Constant Rate Model, the $\beta_0$-curves all start at $500$ because all bubbles are born simultaneously. Looking at the top right panel, it appears as if the patch stage is absent in the Clustered Model. At the height of the patch epoch, there are only $\beta_2=7.0$ neutral patches on average, compared to $\beta_2=48.0$ patches in the Poisson Model. This is because during the patch epoch, the clustered regions are completely filled up and lack any tunnels or patches. The only remaining patches are the huge empty bubble-less regions, which are few in number but large in size. This is obvious when we consider the persistence diagrams for patches in the second column.

In the bottom right panel, we see that the Anti-Clustered Model has a very long pre-overlap stage during which the number $\beta_0$ of ionised regions plateaus. Because the minimum bubble separation $\lambda$ was set rather high, most centres have a closest neighbour at a distance of roughly $\lambda$. Therefore, the bubble network goes through a swift phase transition at $t=\tfrac{1}{2}\lambda=0.047$, when the bubbles start to overlap. We also see that the Anti-Clustered Model has a more significant patch epoch and a brief but extreme filament epoch. The number of tunnels almost reaches 500, nicely adhering to the crystalline expectation.

Looking at the persistence diagrams in the first two columns, we see that although the Clustered Model has fewer higher-dimensional structures, they are far more persistent and appear over a much wider time interval. For the Anti-Clustered Model, we see that there are many more tunnels and patches, but they exist only during a very short period of time. Again, we find that the apparent intensities of the tunnel and patch epochs in the Betti diagrams are deceiving: the Clustered Model does have a patch epoch, but there are fewer yet more significant patches. The opposite is true for the Anti-Clustered Model. The Poisson Model is once more an intermediate case.

The persistence diagrams of the Clustered Model in the top row show additional structure. Observe that there are two distinct generations of features. This is a reflection of the fractal-like multi-scale topology produced by the Neyman-Scott process. The small-scale low-persistence features correspond to structure that emerges early on within clusters. The large-scale high-persistence generation consists of global features that arise when clusters merge with clusters.

\subsection{Spatial filtrations}

We now consider filtrations of bubble networks with a given size distribution. These are spatial filtrations, which means that the interpretation is somewhat different from the temporal filtration described above. The filtration parameter is the scale $\alpha$. At $\alpha=0$, the bubble network has exactly the specified size distribution. Any features that exist at this scale are present in the bubble network. However, we can bring to light additional structure by inflating ($\alpha>0$) or deflating ($\alpha<0$) the bubbles according to equation \eqref{eq:radius_func}. See also figure \ref{fig:log_normal_alpha_slices}.

Let $R_\text{max}$ be the radius of the largest bubble. Starting at any $\alpha<-R_\text{max}$, the filtration is empty. A bubble with radius $R$ enters the filtration at $\alpha=-R$, increasing the number of ionised regions $\beta_0$ by 1. If two bubbles in the network overlap, they must have merged at some $\alpha<0$, at which point $\beta_0$ decreased by 1. The bubble size distribution and merger history is thus directly reflected in the homology for $\alpha<0$. Looking at the homology for $\alpha>0$ allows us to determine the significance of features that exist at $\alpha=0$. For example, features with small $\alpha_\text{death}$ are ephemeral and more likely to be noise. Finally, there are features that exist only over a range of positive scales (i.e. $\alpha_\text{born}>0$). These offer a glimpse of the future of the bubble network. However, because ionisation bubbles do not grow according to equation \eqref{eq:radius_func} in practice, we only get a distorted picture of the future.

\subsubsection{Bubble size distribution}\label{sec:sizes}

To illustrate these ideas, we consider three models with a log-normal size distribution with mean $\mu$ and standard deviation $\sigma$. The average bubble radius is
\begin{align*}
\langle R\rangle = e^{\mu+\sigma^2/2}.
\end{align*}

\noindent
The results for $N=500$ bubbles, again averaged over 10 realisations, are shown in figure \ref{fig:log_normal_results}. The bubbles are smallest in the top row ($\mu=-4.0$, $\sigma=0.50$), largest in the bottom row ($\mu=-2.5$, $\sigma=0.05$), with the middle row being an in-between case ($\mu=3.0$, $\sigma=0.10$). 

The first two columns show the persistence fields for ionised bubbles ($\beta_0$) and neutral filaments ($\beta_1$). We have divided the persistence fields into quadrants. Features in quadrant II ($\alpha_\text{birth}\leq0$, $\alpha_\text{death}>0$) are present in the bubble network. Features in quadrant I exist only over positive scales. The Betti curves in the third column indicate the total numbers of features alive at every scale. The homology of the bubble network can be read off by looking at the intersections of the Betti curves with the line $\alpha=0$. 

We have overlaid the log-normal probability density functions on the $\beta_0$ persistence fields. By construction, the scales $\alpha_\text{birth}$ at which bubbles are born follow the log-normal distribution. In the top row, we see that the largest bubbles die at large scales. But many of the smallest bubbles die at small scales. This is because of the elder rule, which states that whenever two features merge, the oldest feature survives. We have also overlaid the cumulative distribution functions $P(R\geq -\alpha)$ on the Betti curves. The $\beta_0$-curve follows the distribution function when $\alpha$ is small, but deviates once bubbles start to merge.

We can identify what stage of the reionisation process the bubble network is currently undergoing by considering the distribution of features over the quadrants:

\begin{enumerate}[label=(\roman*), wide=0pt, widest=99,leftmargin=\parindent, labelsep=*]
	\item When the bubbles are smallest (top), almost all features in the $\beta_0$-field are in quadrant II and all features in the $\beta_1$-field are in quadrant I. We conclude that the topology is dominated by separated islands of ionised material. The bubble network is in the pre-overlap stage.
	\item In the second row, about a third of the 0-dimensional features are in quadrant II and two thirds are in quadrant III. Most of the tunnels are in quadrant I, although a few are in quadrants II and III. The bubble network is therefore in the overlap stage, and entering the filament stage.
	\item Finally, when the bubbles are largest (bottom), all 0-dimensional features are in quadrant III. This means that they have all merged into a single connected component. A preponderance of the tunnels are in quadrant II and therefore alive in the bubble network. We are in the filament stage.
\end{enumerate}

In each case, a quick glance at the bubble network slices shown in figure \ref{fig:log_normal_slices} confirms the picture sketched by the division of features over the quadrants. Overall, the Betti curves and persistence fields most resemble the Poisson Model. This is not surprising, because the source locations were generated with a Poisson point process. Finally, note that the $\beta_1$-fields look like translated and distorted copies of each other. The reason that the fields are not just translated copies is because of the non-linear bubble scaling $\sim\sqrt{R^2+\alpha^2}$.

\section{Discussion}\label{sec:conclusion}

The formalism presented here provides a substantial deepening of our understanding of the topology of \HII\, regions during the Epoch of Reionisation. Homology allows us to characterise the topology of the ionisation bubble network in terms of its components (ionised regions), tunnels (enclosed neutral filaments), and cavities (neutral patches), collectively called topological features. Persistence is a measure of the significance of a feature. We have shown that the persistence of a feature can be variously interpreted as its lifetime, significance, or ionisation state. Persistent homology provides us with quantitative measures that are more general than other commonly used measures such as the Euler characteristic, Minkowski functionals, and the bubble size distribution. 

We use the tool of $\alpha$-shapes, borrowed from computational topology, to model the ionisation bubble network. Together with persistent homology, $\alpha$-shapes are ideally suited to study every stage of the reionisation process. Starting at the pre-overlap and overlap stages, we can follow the number of distinct ionised regions as bubbles arise and subsequently merge. During the later stages, homology allows us to understand the topology of the bubble network as a large fractal-like structure, pierced by neutral filaments and enclosing patches of neutral hydrogen. The topology of the bubble network depends on the underlying physics through an interaction of the spatial distribution of the ionising sources with the size distribution of the surrounding $\HII$ regions.

This work is a stepping stone for a number of further studies on the persistent topology of reionisation. In an upcoming paper, we apply our methods to a physical model of reionisation. Ultimately, the goal is to study the topology of reionisation using 21-cm observations. As illustrated with the phenomenological models in this paper, what is needed is a specification of the spatial and size distribution of $\HII$ regions. A first step will be to study the viability of sufficiently constraining these properties with upcoming observations. Further statistical analysis will be necessary to determine the requirements on future experiments that would definitively allow a homological study of reionisation. However, the statistics of persistence diagrams has only recently been put on firm footing, so more work is needed on this front. Finally, we have proposed a more general filtration method that relaxes the $\alpha$-shape assumptions. This method would enable us to study the evolving topology of a fully dynamic ionisation fraction field, given a set of 3D measurements of the ionisation field at multiple redshifts.

\section*{Acknowledgements}

RvdW is grateful for numerous useful, instructive and insightful discussions with Gert Vegter, Bernard Jones, Job Feldbrugge, Garrelt Mellema and Keimpe Nevenzeel. WE similarly thanks Martijn Oei and Kees Elbers for helpful discussions.





\bibliographystyle{mnras}
\bibliography{main}



\appendix

\label{lastpage}
\end{document}